\documentclass[pra,superscriptaddress,floatfix,twocolumn,amsfonts,amsmath,amssymb]{revtex4}
\usepackage[english]{babel}
\usepackage{latexsym}
\usepackage{graphics}
\usepackage{graphicx}
\usepackage{epsfig}
\usepackage{color}
\usepackage{bm}
\usepackage{amsmath}
\usepackage{amssymb}
\usepackage{amsthm}
\usepackage{dcolumn}
\usepackage{bm}
\usepackage{float}
\usepackage{hyperref}
\usepackage{color}
\usepackage{epstopdf}
\usepackage{cleveref}
\usepackage[svgnames]{xcolor}
\hypersetup{hidelinks,colorlinks=true,allcolors=DarkBlue}

\theoremstyle{remark}

\begin{document}

\preprint{APS/123-QED}

\title{Revealing quantum chaos with machine learning}

\author{Y.A. Kharkov} 
\affiliation{Russian Quantum Center, Skolkovo, Moscow 143025, Russia}
\affiliation{School of Physics, University of New South Wales, Sydney 2052, Australia}
\affiliation{Joint Center for Quantum Information and Computer Science, NIST/University of Maryland, College Park, Maryland 20742, USA}

\author{V.E. Sotskov}
\affiliation{Russian Quantum Center, Skolkovo, Moscow 143025, Russia}

\author{A.A. Karazeev}
\affiliation{Russian Quantum Center, Skolkovo, Moscow 143025, Russia}
\affiliation{Moscow Institute of Physics and Technology, Dolgoprudny, Moscow Region 141700, Russia} 
\affiliation{QuTech, Delft Technical University, 2600 GA Delft, The Netherlands}

\author{E.O. Kiktenko}
\affiliation{Russian Quantum Center, Skolkovo, Moscow 143025, Russia}
\affiliation{Steklov Mathematical Institute of Russian Academy of Sciences, Moscow 119991, Russia}
\affiliation{NTI Center for Quantum Communications, National University of Science and Technology MISIS, Moscow 119049, Russia}

\author{A.K. Fedorov}
\affiliation{Russian Quantum Center, Skolkovo, Moscow 143025, Russia}
\affiliation{Moscow Institute of Physics and Technology, Dolgoprudny, Moscow Region 141700, Russia} 

\date{\today}
\begin{abstract}
Understanding properties of quantum matter is an outstanding challenge in science. In this paper, we demonstrate how machine-learning methods can be successfully applied for the classification of various regimes in single-particle and many-body systems. 
We realize neural network algorithms that perform a classification between regular and chaotic behavior in quantum billiard models with remarkably high accuracy. 
We use the variational autoencoder for autosupervised classification of regular/chaotic wave functions, as well as demonstrating that variational autoencoders could be used as a tool for detection of anomalous quantum states, such as quantum scars. 
By taking this method further, we show that machine learning techniques allow us to pin down the transition from integrability to many-body quantum chaos in Heisenberg XXZ spin chains. 
For both cases, we confirm the existence of universal W shapes that characterize the transition. 
Our results pave the way for exploring the power of machine learning tools for revealing exotic phenomena in quantum many-body systems.
\end{abstract}

\maketitle

\section{Introduction}
The idea of combining machine learning methods~\cite{LeCun2015} with quantum physics has stimulated an intensive research activity~\cite{Biamonte2017}.
The scope so far includes identification of quantum phases of matter and detecting phase transitions~\cite{Wang2016, Broecker2016,Wetzel2017,Melko2017,Schindler2017,Chng2017,Nieuwenburg2017,Ringel2018,Beach2018,Greitemann2018,Knap2019},  
representation of states of quantum many-body systems~\cite{Troyer2017,Glasser2018,Lu2018,Troyer2018}, and machine-learning-based analysis of experimental data~\cite{Troyer2018,Zhang2018,Sriarunothai2018}.

Remarkable progress on building large-scale quantum simulators~\cite{Monroe2013,Rey2017,Lukin2017,Monroe2018} 
has opened fascinating prospects for studying traditionally challenging problems of complex quantum systems, such as investigation of quantum critical dynamics and quantum chaos~\cite{Polkovnikov2016}.
Quantum systems with chaotic behaviour are of great interest particularly in the view of a possibility to explore many-body quantum scars~\cite{Papic2018,Ho2019}, which can be  compatible with long-lived states.
A standard criterion for the separation between regular and chaotic regimes uses the nearest-neighbor (NN) energy level statistics~\cite{Berry1977, Bohigas1984}: 
Poisson and Wigner-Dyson distributions correspond to integrable and chaotic systems, respectively. 
However, the energy level statistics of highly excited states is not always  accessible in experiments with well-controlled quantum systems.

From the machine learning perspective, an interesting problem is to understand whether it is possible to distinguish between regular and chaotic behavior 
based on experimentally accessible quantities such as data from projective measurements.
This question can be further extended to a possibility to detect anomalies in experimental data, such as quantum scars.

\begin{figure}[h!]
\includegraphics[width=0.49\textwidth]{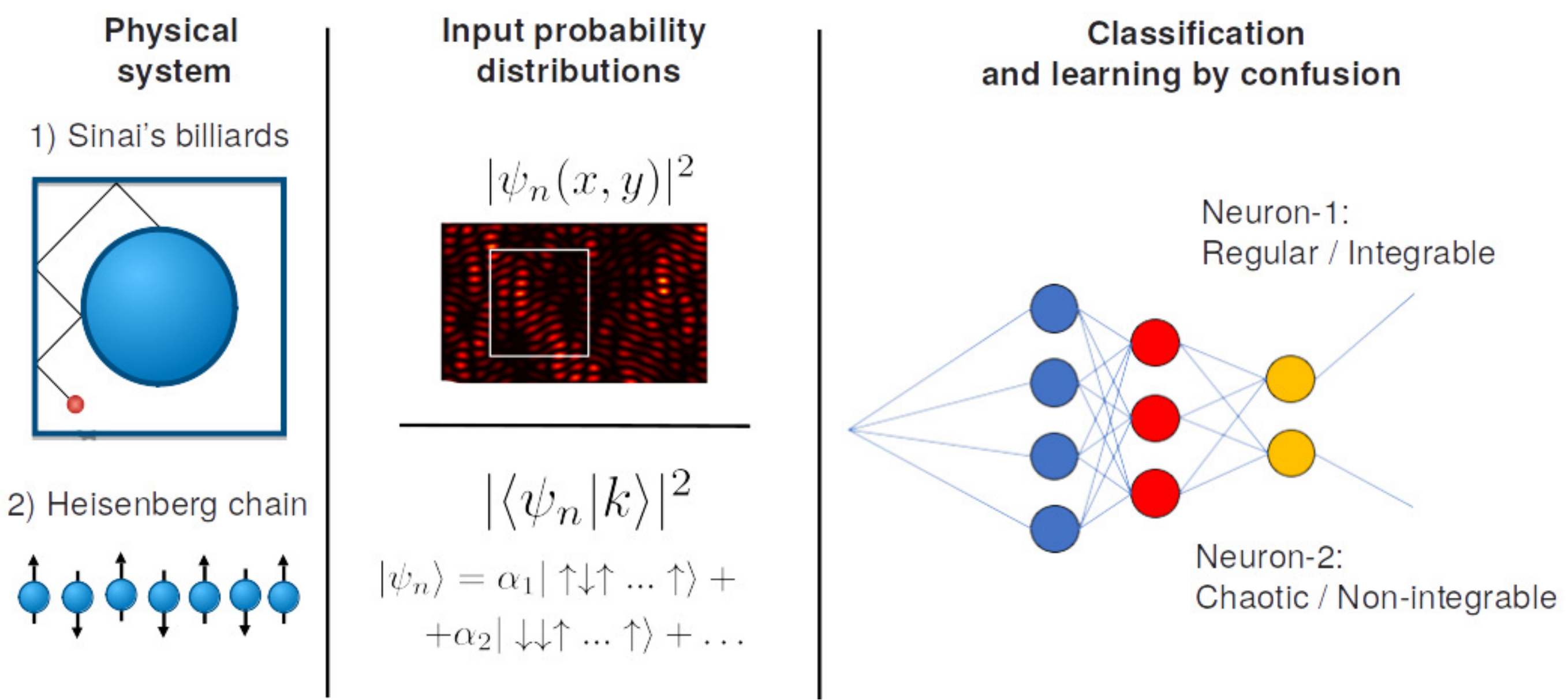}
\vskip -4mm
\caption{Neural network approach for identifying a transition between chaotic and regular states in quantum billiards and Heisenberg spin chains. 
The input data contains probability distribution in the configuration space, the two neuron activation functions are used for the identification of the two regimes.}
\label{fig:NN}
\end{figure}

In this paper, we realize machine learning algorithms to perform a classification between regular and chaotic states in single-particle and many-body systems.  
The input data contains probability density function representing configurations of excited states and the output is provided by two neurons, which distinguishes between integrable and chaotic classes, see Fig. \ref{fig:NN}.
In the single-particle case, we consider paradigmatically important models of quantum billiards.
We apply an extension of a semisupervised `learning by confusion'  scheme~\cite{Nieuwenburg2017} in order to detect the integrability/chaos transition and to evaluate a critical critical region.
We also use a clusterization technique based on variatonal autoencoder (VAE) for machine learning of the transition to quantum chaos and for
detection of quantum scars.   
The supervised approach is then extended in order to study the transition in Heisenberg XXZ spin-1/2 chain in the presence of additional interactions that break integrability.
In our work, regular/chaos transitions are identified with the classification accuracy up to $99\%$.
We show that our results based on the machine learning approach are in a good agreement with the analysis of level spacing distributions.  

The confusion scheme is based on the assumption that the critical point $\lambda^{c}$ exists within a given parameter range $(a, b)$, so that the data could be classified into two classes. 
Further, a trial critical point $\lambda^{c}$ is proposed and all the data with parameters below $\lambda^{c}$ is labelled as 0, and above $\lambda^{c}$ as 1. 
Neural network is then trained on the entire dataset for all values of $\lambda^{c}$, chosen from the range $(a, b)$ with a predefined step. 
This method results in a universal W-like performance curve~\cite{Nieuwenburg2017}.
The `learning by confusion'  scheme has been used for the study of many-body localization--delocalization transition~\cite{Gornyi2018}, 2D percolation and Ising models~\cite{Zhao2019}, 
critical behavior of the two-color Ashkin-Teller model, the XY model, and the eight-state clock model~\cite{Lee2019},
and exploring topological states~\cite{Granath2019}.

To address the problem of revealing the transition between regular and chaotic behaviour, 
we realized an extension of the `learning by confusion'  scheme.
At the first stage, we train the network to distinguish states belonging to the extreme cases of regular ($\lambda = 0$) and chaotic ($\lambda\sim1$) regimes, where $\lambda$ is the chaoticity parameter. 
By analyzing neural network outputs, we determine the critical domain where the neural network predicts a transition between the two regimes. 
At the second stage, we perform the standard `learning by confusion' protocol and we refer the middle peak on W-like performance curves of the neural network as the transition point~\cite{Nieuwenburg2017}. 

\begin{figure}[h!]
\includegraphics[width=.48\textwidth]{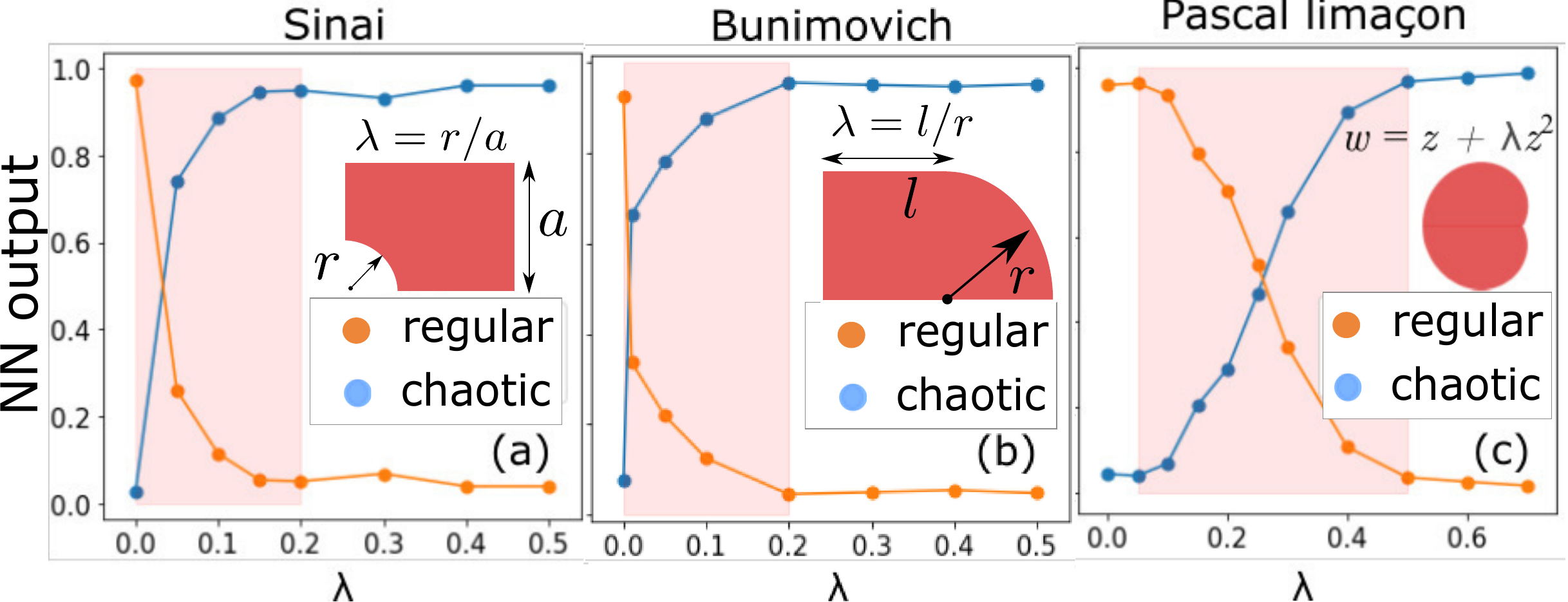}
\vskip -2mm
\caption
{Convolutional neural network outputs for (a) Sinai billiard, (b) Bunimovich stadium, and (c) Pascal's lima\c{c}on as functions of the chaoticity parameter $\lambda$ characterizing billiard's boundary shape.
The highlighted critical region corresponds to the regions of ``uncertainty'' in neuron network output activation curves.}
\label{fig:activ_bill}
\end{figure}

The paper is organized as follows. In Section  \ref{sec:bills} we describe our machine learning  approach for classifications of chaotic/integrable wavefunctions in quantum billiards. 
We describe our supervised learning methodology and present results of neural network-based classification of chaotic/integrable wavefunctions as a function of chaoticity parameter $\lambda$ for three types of quantum billiards: 
Sinai, Bunimovich stadium and Pascal billiards. 
In Section \ref{sec:vae} we apply autosupervised machine learning techinique using variational autoencoder (VAE) for clusterization analysis of quantum states in billiards. 
In addition we explore applications of VAE for anomaly detection of quantum scars and the potential of VAE for generative modeling of regular/chaotic wavefunctions in quantum billiards.
In Section \ref{sec:xxz} we apply supervised learning for detection of transition from integrability to quantum chaos in XXZ spin chains in the presence of integrability breaking interactions: 
next nearest neighbor spin-spin interaction and a local magnetic field.
We draw conclusion in Section \ref{sec:concl}.
Technical details on machine learning approaches and datasets preparation are presented in Appendixes.

\section{Quantum billiards}\label{sec:bills}

Quantum billiards are among the simplest models exhibiting quantum chaos. 
The transition from regular to chaotic behaviour in quantum billiards, which is controlled by the shape of the billiard boundary, has been intensively studied for decades~\cite{Jain2017}. 
Quantum billiards have been realized in various experimental setups including microwave cavities~\cite{Sridhar1991}, ultracold atoms~\cite{Raizen2001}, and graphene quantum dots~\cite{Geim2008}.
Quantum scars~\cite{Heller1993}, which are regions with enhanced amplitude of the wave function in the vicinity of unstable classical periodic trajectories, 
is the hallmark of quantum chaos. 
Quantum scars are of a great interest in quantum billiards~\cite{Heller1993,Tao2008} and their many-body analogs have recently been studied~\cite{Papic2018,Ho2019}.

We consider three standard types of two-dimensional quantum billiards: Sinai billiard, Bunimovich stadium, and Pascal's lima\c{c}on (Robnik) billiard. 
We define a dimensionless parameter of chaoticity $\lambda$ for each billiard type, where it determines the billiard shape. 
In Sinai billiard the chaoticity parameter is controlled by the ratio of the inner circle radius to the width/height of the external rectangle, so  $\lambda = r/a$.
In the case of Bunimovich stadium the parameter is $\lambda = l/r$ and in the 
Pascal's lima\c{c}on billiard shape is defined via the conformal map on the complex plane $D(w):\{w=z+\lambda z^2\}$, where $|z|\leq1$. At the limit of $\lambda \rightarrow 0$ these billiards have regular shapes and therefore are integrable. 
Varying the parameter $\lambda$ allows one to trace out a continuous transition from integrability to quantum chaos.

We use a supervised learning approach for revealing chaotic/regular transitions in quantum billiard models.
We train a binary classifier based on convolutional neural network (CNN) using real space images of the probability density function (PDF) $|\psi_n(x,y)|^2$. 
The training dataset consist of randomly sampled snapshots of the PDF in fragments excluding the billiard's boundary in the regions of interest.  
The wave functions $\psi_n(x,y)$ are obtained from the numerical solution of the stationary Schr{\"o}dinger equation for the corresponding billiard type (for details  see Appendixes~\ref{app:num_schro}, \ref{app:dataset_bill}).
Since the information about the transition from the regular to chaotic regimes is mostly represented in the properties of highly excited states, we use wave functions with sufficiently large values of $n$ in our dataset.

The snapshots corresponding to $\lambda = 0$ we label as ``regular'' (class 1), and snapshots corresponding to $\lambda\sim 1$ we label as ``chaotic'' (class 2). 
The activation function of the two neurons in the last layer allows classifying between chaotic/regular snapshots in the test dataset with a high accuracy. 
CNN performance curves for each of the three billiard types for different values of $\lambda$ show that the CNN algorithm is able to learn the difference between regular and chaotic wave functions and reveals the existence of the transition region (see Fig.~\ref{fig:activ_bill}). 
The CNN confidence for the binary classification $>95\%$ for $\lambda$ away from the critical region. 
The critical region determined by the CNN is highlighted in red in Fig.~\ref{fig:activ_bill}.  
In Sinai and Bunimovich billiards the critical region detected by the CNN algorithm is $0<\lambda_{c}<0.2$. 
The detected critical region for the Pascal billiard is $0.05<\lambda<0.5$. 
The boundaries of the critical regions provided by the CNN classifier are in a good agreement with the ones obtained from the analysis of the energy levels spacing statistics, see Appendix  \ref{app:en}.

The critical region can be analyzed in more details within the `learning by confusion'  scheme~\cite{Nieuwenburg2017} by performing a dynamical reassignment of the class labels with respect to a given value of $\lambda$. 
We note that a precise definition of the transition point $\lambda_c$ is somewhat ambiguous and depends on  selected criteria, because all observables have a smooth dependence on the parameter $\lambda$. 
Therefore, in our approach we only estimate the location of a characteristic critical point $\lambda_c$, separating regular and chaotic regimes.
The estimated position of the critical point is $\lambda_c\approx 0.1$ in Sinai billiard and $\lambda_c\approx 0.2$ in Pascal lima\c{c}on billiard. 
The location of the critical point $\lambda_c$ in Pascal's billiard agrees with Ref. \cite{Prosen1993}. 
We note that the analysis of the chaotic/regular transition for the Bunimovich stadium is challenging due to its extreme sensitivity to the variation of the chaoticity parameter $\lambda$ (see Ref.~\cite{Tao2008}).

One of the key features that allows us to perform machine learning of the regular-to-chaos transition is the difference in statistical properties of $|\psi_n|^2$ in theses two regimes.
While in the chaotic case the wave functions have Gaussian statistics, in regular case the probability distribution is non-universal and has a power-law singularity at small values of $\psi_n$~\cite{Beugeling2017}.

The standard approach to identify a transition from an integrability to a quantum chaos is based on the comparison of the energy level spacing statistics with the Poisson distribution and the Wigner-Dyson distributions. 
In order to characterize a ``degree of chaoticity'' of the system one can use the average ratio of consecutive level spacings $\langle r \rangle$, 
where $r = \min(\Delta E_{n+1}, \Delta E_n)/\max(\Delta E_{n+1}, \Delta E_n)$ and $\Delta E_n = E_{n}-E_{n-1}$~\cite{Atas2013}.
Here we introduce a different measure based on the Kullback-Leibler (KL) divergence, defined as follows:
\begin{eqnarray}\label{eq:KL}
	D_{KL} (P_\lambda||P') = \int_{0}^{\infty} P_\lambda(s)\log\frac{P_\lambda(s)}{P'(s)}ds,
\end{eqnarray}
where $P_\lambda(s)$ is the level spacing distribution for a given value of $\lambda$,
and $P'(s)$ is the Wigner-Dyson or Poisson distribution:
$P'_{Pois} = e^{-s}$, $P'_{WD} = \frac{\pi}{2}\,s \,\exp\left(- \frac{\pi}{4} s^2 \right)$.
Here $s$ is the unfolded nearest neighbour energy level spacing. 

In the critical region between regular and chaotic regimes the energy spacings distribution is neither the Poisson nor the Wigner-Dyson. 
There exists a point $\lambda_c$ when $P_{\lambda_c}$ is equidistant from both Poisson and Wigner-Dyson distributions within the KL metric, $D(P_{\lambda_c}||P'_{Pois}) = D(P_{\lambda_c}||P'_{WD})$, which we refer as a ``critical point''.
The critical points predicted by the confusion scheme and KL divergence curves are in a good agreement.
We note that the confusion scheme uses experimentally accessible quantities, whether energy levels statistics from experimental data is hardly accessible in condensed matter and atomic simulator experiments.

\begin{figure}[h!]
\includegraphics[width=0.45\textwidth]{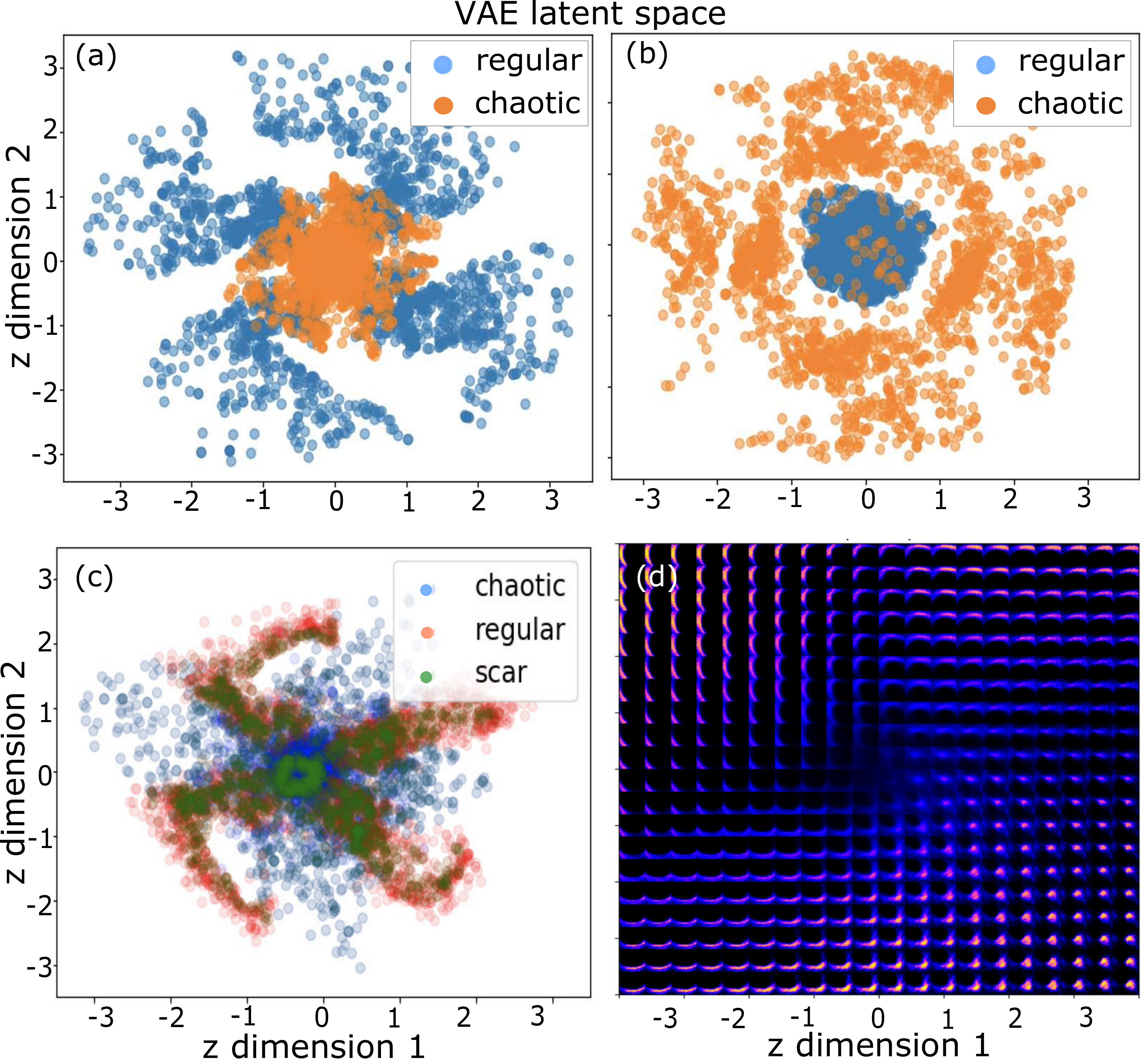}
\vskip -4mm
\caption{Autosupervised learning of regular and chaotic states in quantum billiards with variational autoencoder (VAE). 
Latent space representation of the wave functions in (a) Bunimovich stadium, (b) Sinai billiard; $z_{1,2}$ are coordinates in the two-dimensional latent space. 
(c) Anomaly detection: latent space representation of wave functions corresponding to regular (red dots, $\lambda=0$) and chaotic (blue dots, $\lambda=0.4$) wave functions  as well as scarred chaotic wave functions (green dots) in Bunimovich billiard. 
(d) VAE as a generative model:  images of wave functions $|\psi|^2$ generated by VAE corresponding to different  position in the latent space variables $(z_1, z_2)$  (Pascal billiard). 
By continuously scanning across two-dimensional latent space VAE performs a smooth interpolation between wave functions from chaotic and regular wave functions.}
\label{fig:vae_bill}
\end{figure}

\begin{figure*}[htbp]
\includegraphics[width=1\linewidth]{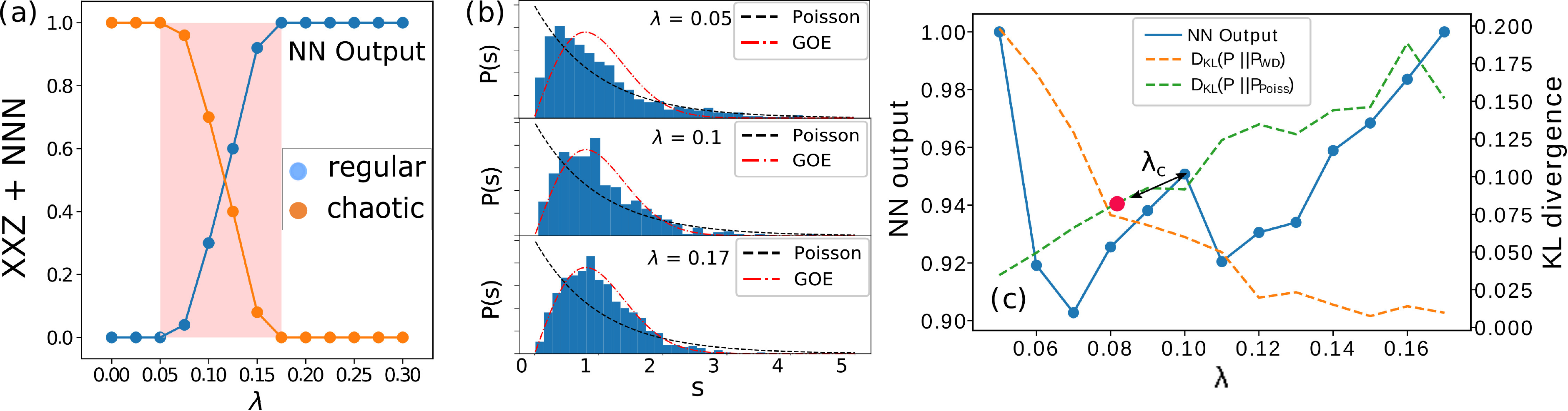}
\vskip -3mm
\caption
{Neural network classification accuracy between integrable and chaotic XXZ spin chains with the next-nearest neighbor interactions for $N=15$ spins  ($J_{zz}= 1$). 
(b) Distribution of energy level spacings and the Poisson/Wigner-Dyson distributions.
Plots correspond to XXZ model . (c) `Learning by confusion' W-like NN performance curve.}
\label{fig:distr}
\end{figure*}

\section{VAE and anomaly detection in quantum billiards}\label{sec:vae}

An alternative approach to differentiate between regular and chaotic regimes is to use auto-supervised machine learning techniques, such as VAE. 
VAEs are generative NN models that are able to directly learn statistical distributions in raw data and can be efficiently used for solving clustering problems~\cite{Kingma2014, Sohn2015}. 
VAE consists of encoding NN, latent space and decoding NN, Fig. \ref{fig:vae_bill}a. 
During the training VAE ``learns'' to reproduce initial data by optimizing the weight in the encoder and decoder NN and parameters in the latent layer. 
Training VAE on the images with  $|\psi_n(x,y)|^2$ corresponding regular ($\lambda=0$) and chaotic ($\lambda\sim 1$) cases and 
by taking samples from the latent space with the dimension 2 results in two clearly separated clusters representing regular and chaotic wave functions. For details on VAE architecture and optimization see Appendix \ref{app:VAE}.

In Figs.~\ref{fig:vae_bill}(a) and ~\ref{fig:vae_bill}(b) we demonstrate latent space representation of wave functions in Bunimovich and Sinai billiards.
The separation in the two clusters shows that VAE is able to learn the difference in the statistical properties of $|\psi_n|^2$ in regular and chaotic billiards. 
Similar approach was used for unsupervised learning of phase transitions~\cite{Wetzel2017}.

In addition to the autosupervised learning of regular/chaotic quantum states, VAE could be used as a tool for anomaly detection in quantum data, in particular identification of  scarred wave functions.  
In this context we use  the term 'anomalous'  to describe a subset of samples with statistical properties drastically different from the statistical properties of the entire dataset. 
Data-driven anomaly detection with VAEs arises in machine learning, data mining and cybersecurity applications~\cite{Golan2018, An2015, Zhang2019}. 
Applications of VAE-based anomaly detection methods  were recently  studied in  the context of classical phase transitions~\cite{Cristoforetti2017} and detection of elementary particles~\cite{cern2018}.
However, potential of anomaly detection methods in quantum systems  has been mostly unexplored.
Anomalous samples could be detected using latent space  representation   $z_{1,2}$ as a set a cloud of points falling outside of the  `chaotic` cluster (for additional details see Appendix~\ref{app:anom}).
Using a pretrained VAE we generate a set of points in the latent space corresponding to the scarred chaotic wave functions, see Fig.~\ref{fig:vae_bill}c.
The `anomalous` cluster representing scarred  wave functions falls outside of the `chaotic' cluster 
and has a large overlap with a  'regular' cluster. This unusual behaviour indicates  similarity  between  scarred wave functions  and wave functions in integrable billiards. 
Interesting extension of this approach could be VAE-based anomaly detection method for identification of quantum many-body scars. 

Another additional feature of VAE is the ability to smoothly interpolate between datasets corresponding to the two classes. 
In Fig.~\ref{fig:vae_bill}d we show  wave functions generated by VAE in Pascal billiard via scanning across the two-dimensional latent space $z_{1,2}$. 
This procedure allows to perform  continuous interpolation between chaotic (center region)  and regular wave functions (outer region).   
VAE-based generative modeling of quantum states could give rise to new approaches in simulations of quantum systems~\cite{Rocchetto2018} as well as for new applications in the context of quantum chaos.
Exploring a full potential of unsupervised machine learning methods for clustering quantum states is beyond the scope of the present paper. 

\section{Detection of quantum chaos in XXZ spin chains}\label{sec:xxz}

While quantum billiards is an instructive example of a single particle quantum chaos, quantum chaotic regimes in many-body systems are more interesting.
Developing machine learning approaches to characterize/classify many-body states in chaotic and integrable regimes using only limited information from measurements is a non-trivial task. 
For example, such techniques can benefit from the analysis of experimental data from quantum simulators~\cite{Monroe2013,Rey2017,Lukin2017,Monroe2018}.
As a prototypical example of a quantum many-body integrable system we consider 1D Heisenberg XXZ spin chain, which is of great interest for realizing models of quantum magnetism using quantum simulators~\cite{Bloch2017}.
Recent experimental advances have opened exciting prospects for exploiting a rich variety of tunable interactions in Rydberg atoms~\cite{Browaeys2016,Lukin2016,Lukin2017,Browaeys2018,Browaeys2018-2} 
and cold polar molecules~\cite{Buchler2012,Ye2012,Rey2013} for engineering of spin Hamiltonians including the XXZ model.

The Hamiltonian of the Heisenberg XXZ model reads:
\begin{eqnarray}
	H_{XXZ} = \sum_{i=1}^{N-1} \left[ J(S_i^{x}S_{i+1}^{x} + S_i^{y}S_{i+1}^{y})+J_{zz}S_i^{z}S_{i+1}^{z} \right], \label{eq:XXZ_Ham}
\end{eqnarray}
where $N$ is the number of spins, $J$ and $J_{zz}$ are the Heisenberg exchange constants and $S_i^{x,y,z}$ are Pauli spin-1/2 operators. 
For simplicity we only consider  antiferromagnetic XXZ model, $J,J_{zz}>0$.
Hereafter we set $J=1$. 
The XXZ model is integrable and exactly solvable by Bethe ansatz~\cite{Buchler2012}, however it can be non-integrable in the presence of additional interactions.

\begin{figure}[h!]
\includegraphics[scale=0.42]{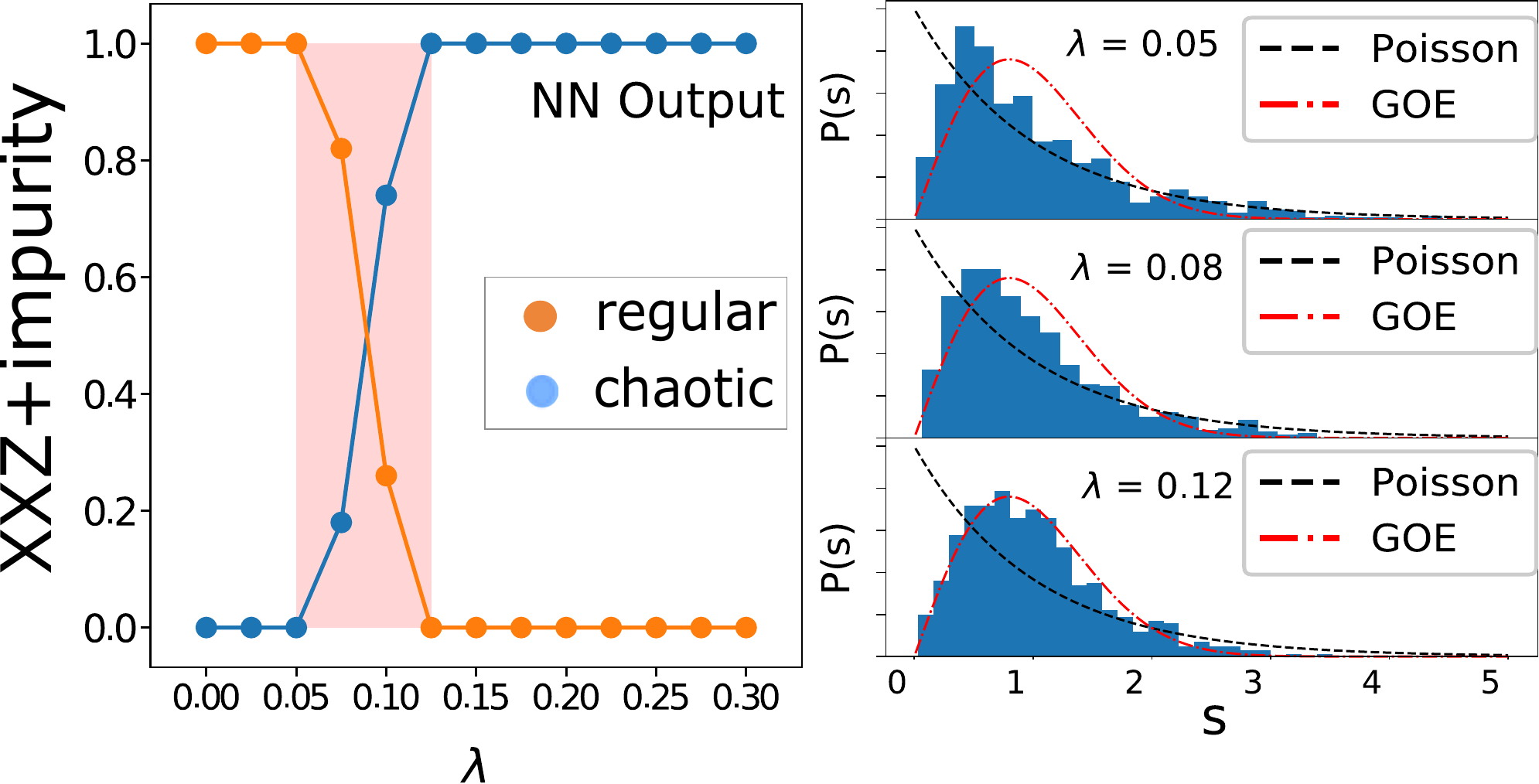}
\caption{(Left panel) NN  classification accuracy  for  chaos-integrability transition in XXZ model in the presence of a local magnetic field (a magnetic impurity) for $N=15$ spins. (Right panel) Energy level spacing distributions for different values of $\lambda$. }\label{fig:fig5_xxz_imp}
\end{figure}

Here we consider two types of perturbations that violate integrability of the XXZ model: (i) antiferromagnetic next-nearest neighbour spin-spin interaction (NNN), (ii) a local static magnetic field acting on a single spin (impurity).
We parametrize  perturbations to the Hamiltonians in the following form:
\begin{eqnarray}\label{eq:H_pert}
\textrm{(i)}:\,\, H' = \lambda \sum_{i=1}^{N-2}S_i^{z}S_{i+2}^{z},\quad
\textrm{(ii):}\,\, H' = \lambda S_{(N+1)/2}^{z}.
\end{eqnarray}
We consider spin chains with an odd number of spins $N$, so that in the case (ii) the local magnetic field is acting on the spin in the middle of the chain, i.e. $i=(N+1)/2$.
The Hamiltonian of the perturbed XXZ model reads:
\begin{eqnarray}
	H = H_{XXZ} + H'.
\end{eqnarray}

We train a multilayer perceptron (MLP) on the dataset containing the probabilities  $|\langle \psi_n | k \rangle|^2$ of the spin configurations in $S_z$ representation ($|k \rangle$ refers to basis states in $S_z$-representation), 
e.g. $|\uparrow \downarrow  \ldots \ \downarrow \rangle$. 
The eigenfunctions $|\psi_n\rangle$ are obtained by exact diagonalization of spin-chain Hamiltonian (for details see Appendix \ref{app:data_xxz}), here we consider system size $N=15$. 
Similarly to the case of quantum billiards, we consider only highly excited states with $n$ corresponding to the levels lying in the middle of the energy spectrum, $E_n\approx 0$. 

To pindown the chaos/integrability transition 
we use a MLP NN, see details in Appendix \ref{app:mlp}.
We evaluate NN classification prediction for the test dataset as a function of $\lambda$, see Fig. \ref{fig:fig5_xxz_imp}a,  the critical region is highlighted with red. 
For XXZ + NNN (Fig.  \ref{fig:distr}) and XXZ + impurity  (Fig. \ref{fig:fig5_xxz_imp}) detected critical regions are $0.05\leqslant\lambda_{c}\leqslant0.175$ and $0.05\leqslant\lambda_{c}\leqslant0.125$ respectively, 
which turn out to be in agreement with level spacing distributions represented in Fig. \ref{fig:distr}b, see Appendix~\ref{app:en} and within the range of values obtained in  previous works \cite{Poilblanc1993,  Gubin2012, Wells2014}.
Within these critical regions `learning by confusion'  
resulted in W-like  performance curves, (see Fig.~\ref{fig:distr}c and Appendix \ref{app:w}), 
 and detected transition points $\lambda_{c} \approx 0.1$ for XXZ + NNN and $\lambda_{c} \approx 0.085$ for XXZ + impurity.  
We note that we have a reasonable agreement with the results based on the KL divergence calculations.

\section{Conclusions}\label{sec:concl}

In summary, we have shown the potential of classical supervised and unsupervised machine learning techniques for classification of regular/chaotic regimes in single-particle and many-body systems.
For quantum billiards and XXZ spin chains we demonstrated that neural networks can serve as a binary classifier to distinguish between the two regimes with remarkably high accuracy.  
We revealed the integrability-chaos critical region purely based on machine learning techniques and located the transition point using `learning by confusion'  approach.
The extension of our work opens a new avenue to study chaotic and integrable regimes and detect quantum anomalies using experimentally accessible data in different many-body quantum systems including atomic simulators. 
Harnessing machine learning methods could open up exciting possibilities for studying exotic many-body phenomena with controlled quantum many-body systems, 
such as many-body localization~\cite{Altshuler2006}, many-body quantum scars~\cite{Papic2018}, and ergodic/non-ergodic phase transitions~\cite{Shlyapnikov2018} and near-critical properties of these systems.

\begin{acknowledgments}
{\it Acknowledgments}.
We are grateful to M.B. Zvonarev and V.V. Vyborova for valuable suggestions. 
We thank G.V. Shlyapnikov, V.I. Yudson, and B.L. Altshuler for fruitful discussions and useful comments.
The initial stage of the work was supported by RFBR (Grant No. 18-37-00096). 
The work on the extension of results on the anomaly detection and applications to many-body systems was supported by RSF (19-71-10092).
\end{acknowledgments}

\newpage

\appendix

\section{Numerical solution of the Schr\"{o}dinger equation for quantum billiards.} \label{app:num_schro}
We solve a stationary Schr\"odinger equation describing a single particle in a quantum billiard with the Dirichlet boundary condition:
 \begin{eqnarray}
 -\frac{\hbar^2}{2m}\nabla^2\psi_n=E_n\psi_n, \quad \psi_n|_{\partial D} = 0, \label{eq:Schr}
 \end{eqnarray}
where $\psi_n(x,y)$ is the wave function and $E_n$ is the energy of a particle in the billiard with the boundary $\partial D$; $\nabla^2 = \partial_{xx}+\partial_{yy}$ is the two-dimensional Laplace operator.  
Hereafter we set the Plank's constant and the mass to unity, $\hbar = m = 1$.
In order to solve Eq.~(\ref{eq:Schr}) for an arbitrary 2D billiard boundary shape we use Matlab PDE toolbox. 
The PDE solver is based on the finite element method with an adaptive triangular mesh for a given boundary geometry. 
In order to reduce computational complexity and to avoid additional complications due to degeneracies of eigenstates, we constrain the eigenfunctions to a specific symmetry (parity) sector. 
We remove degeneracies by considering the lowest symmetry segments of billiards.
In the case of the Bunimovich stadium we consider a quarter of the billiard [see inset of Fig.~1(b) in the main text]. 
For the Sinai billiard we consider a boundary with the incommensurate ratio of vertical and horizontal dimensions of the external rectangle, $a_x/a_y = \sqrt{5}/2$ (we denote $a \equiv a_x$ in the main text). 
In the case of the Pascal lima\c{c}on billiard, the degeneracy is lifted when considering only the upper part of the billiard $Re(z)\geq0$.

\section{Dataset preparation  and CNN  for quantum billiards \label{app:dataset_bill}} 

Wave functions $\psi_n(x,y)$ obtained from numerical solution of the Schr\"{o}dinger equation are converted into images of PDFs $|\psi_n(x,y)|^2$. 
From original images with $\sim 500\times 500$ pixels we randomly select square fragments (region of interest) which exclude the billiard boundary, $\sim 300\times 300$ pixels. 
In order to reduce the size of the images we perform a coarse graining (downsampling) to images with dimensions $36\times36$. 
The dataset for each billiard type contains wave functions corresponding to high energy states, $470 \leq n < 500$.
In order to increase amount of images in the dataset we perform an augmentation of the dataset by adding horizontal and vertical reflections, 
discrete rotations by angles $\alpha=k\pi/2$ and rotations by random angles from the uniform distribution $\alpha\in[-25^o, 25^o]$. 
The total number of images in the resulting dataset for each billiard type and each value of $\lambda$ is $M = 4000$. 
The trial samples from the dataset for the Bunimovich billiard are shown in Fig.~\ref{fig:bill_samples}.

\begin{widetext}
\begin{center}
\begin{figure}[h!]
\textsl{} \includegraphics[scale = 0.35]{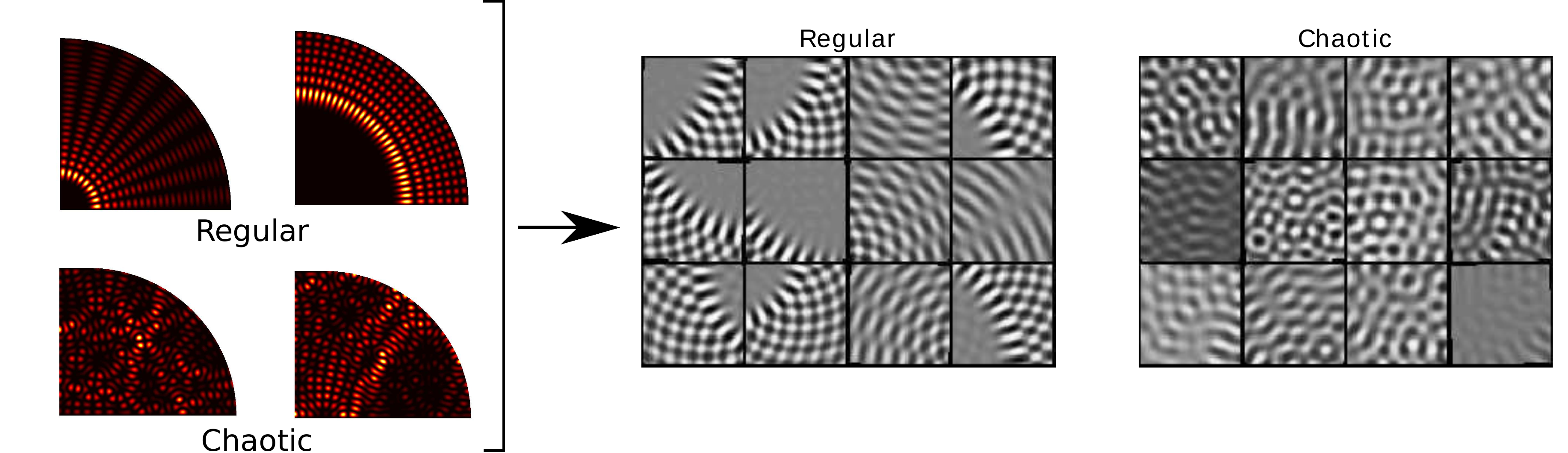}
	\vskip-4mm
	\caption{Sample images of $|\psi(x, y)|^{2}$ in the dataset for Bunimovich billiard. Regular case ($\lambda=0$) and chaotic case ($\lambda=l/r=0.2$).}
	\label{fig:bill_samples}
\end{figure}

\begin{figure}[h!]
	\includegraphics[scale=0.7]{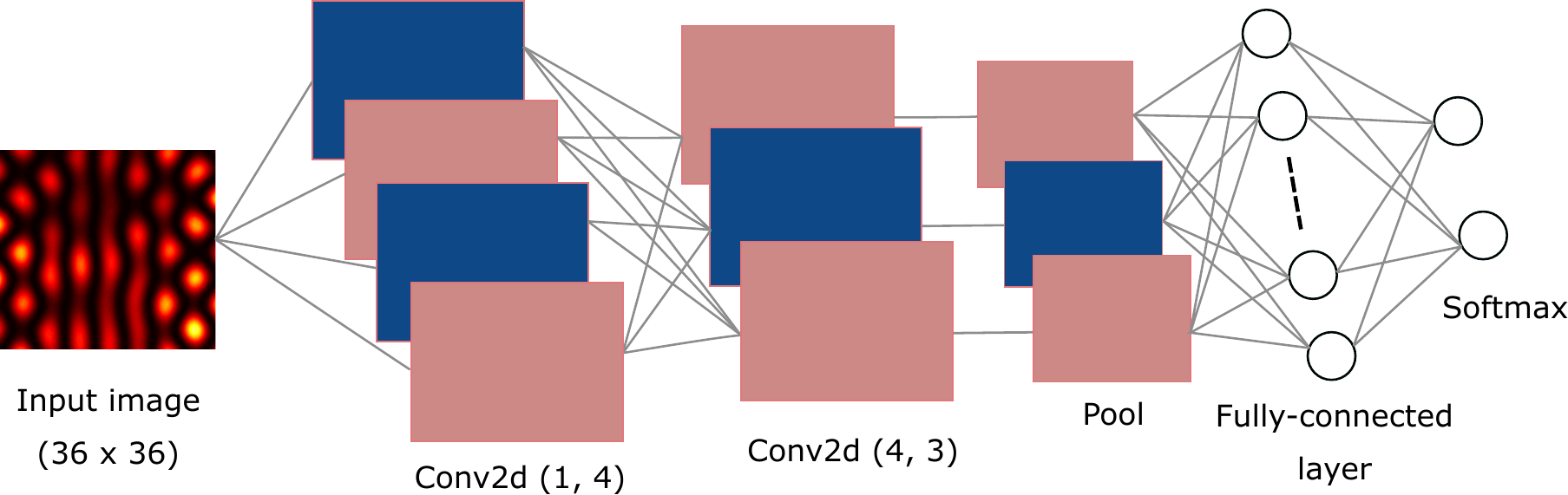}
	\vskip-2mm
	\caption{CNN used for recognizing chaotic regimes in quantum billiards.}
	\label{fig:bill_cnn}
\end{figure}
\end{center}
\end{widetext}

The training dataset consists of labeled images from the class 1 (regular, $\lambda=0$) and class 2 (chaotic, $\lambda = \lambda_0$). 
The value of $\lambda_0$ we independently choose for each billiard type: Sinai - $\lambda_0 = 0.4$, Bunimovich - $\lambda_0=0.2$, Pascal  - $\lambda_0 = 0.8$. 
In order to check that at $\lambda = \lambda_0$ the system is in the chaotic regime we compare the energy level spacing distribution with the Wigner-Dyson distribution. 
As long as the value of $\lambda_0$ is much greater than the critical $\lambda$, $\lambda_0\gg \lambda_c$, the NN activations curves remain practically unchanged (see Fig.~1 in the main text). 

The training and test dataset are split in the proportion $70\%/30\%$. The test set for each billiard type consists of images for several values of $\lambda$ (including values of $\lambda$ not present in the training dataset), 
evaluation of the NN output for the sample images from the test dataset for each value of $\lambda$ results in the NN prediction curves presented in Fig.~1 in the main text.

A CNN consists of two convolutional layers followed by pooling, fully connected and final softmax layers. 
The output from the second convolutional layer is  subject to dropout regularization and batch normalization. 
The cost function for the binary classifier is the cross-entropy and 
the neuron activation function is ReLU. 
The scheme of the CNN architecture is presented in Fig.~\ref{fig:bill_cnn}.
The weights in the CNN are optimized with the use of the Adam optimizer.
The batch size is 60, the number of training epochs is of about $500$, the learning rate is $\alpha = 5\times 10^{-4}$.

\section{Energy level spacing statistics in quantum billiards}\label{app:en}
\begin{widetext}

\begin{figure}[htb!]\label{fig:q_bil_hist}
\includegraphics[scale=0.75]{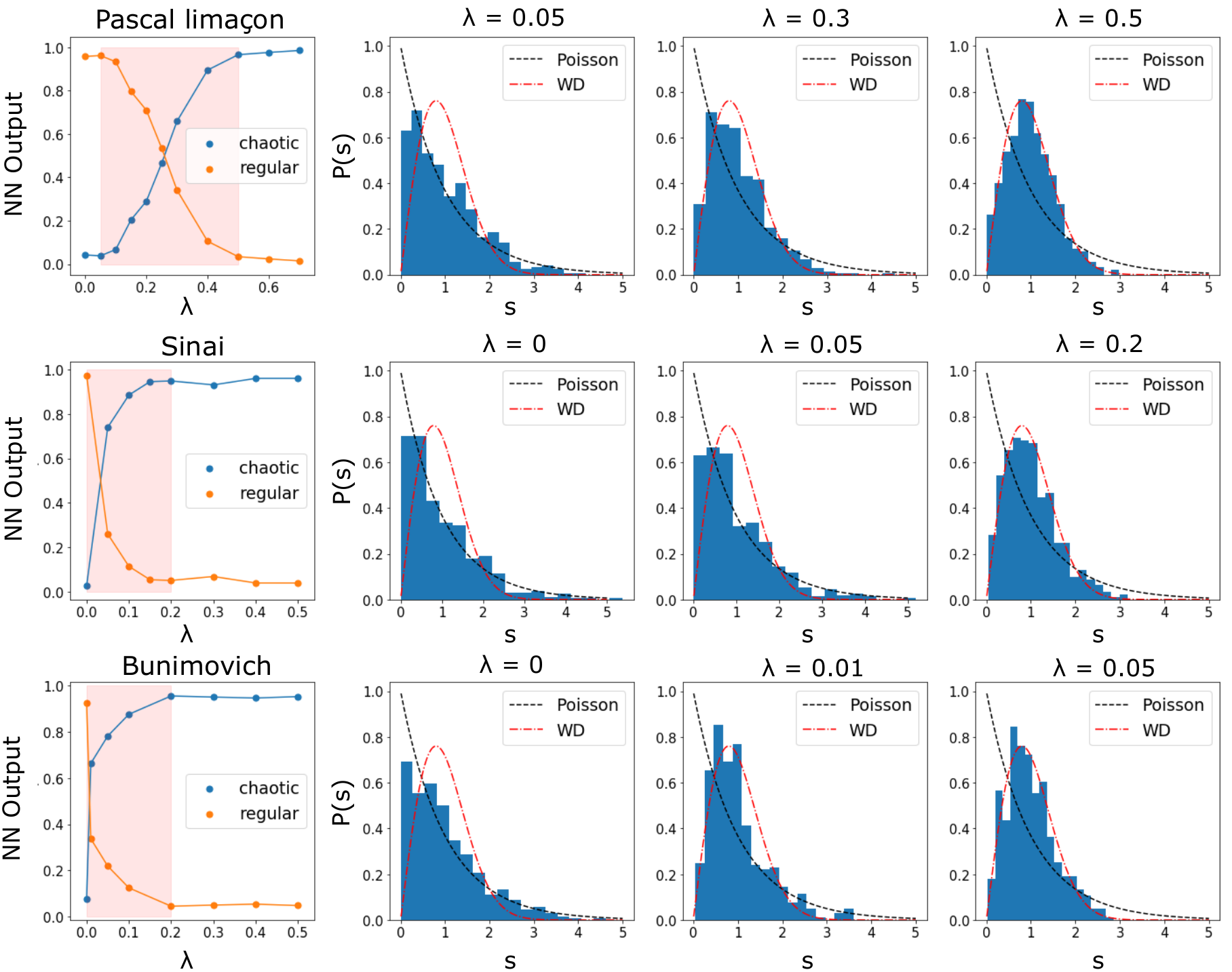}
\caption{Left column: The CNN activation functions (Fig. 1 from the main text). 
The histograms show the energy level spacing distributions (lowest 500 energy levels). 
In order to compare NN prediction for the regular-to-chaos critical region we compare the energy level spacing distribution with the standard Poisson/GOE distributions.}
\end{figure}

\end{widetext}

We validate results of NN classification prediction in quantum billiards (Pascal lima\c{c}on, Sinai and Bunimovich billiards) by comparing NN predictions with the energy level spacing distributions, see Fig. \ref{fig:q_bil_hist}.
In the regular case the energy level spacing distribution $P(s)$ is close to the Poisson distribution (black dashed line), in the chaotic case $P(s)$ is approaching Wigner-Dyson distribution (red dashed dotted line).

\section{Unsupervised learning with VAE}\label{app:VAE}

\vspace{.5cm}
\begin{figure}[htb!]
\includegraphics[scale=0.4]{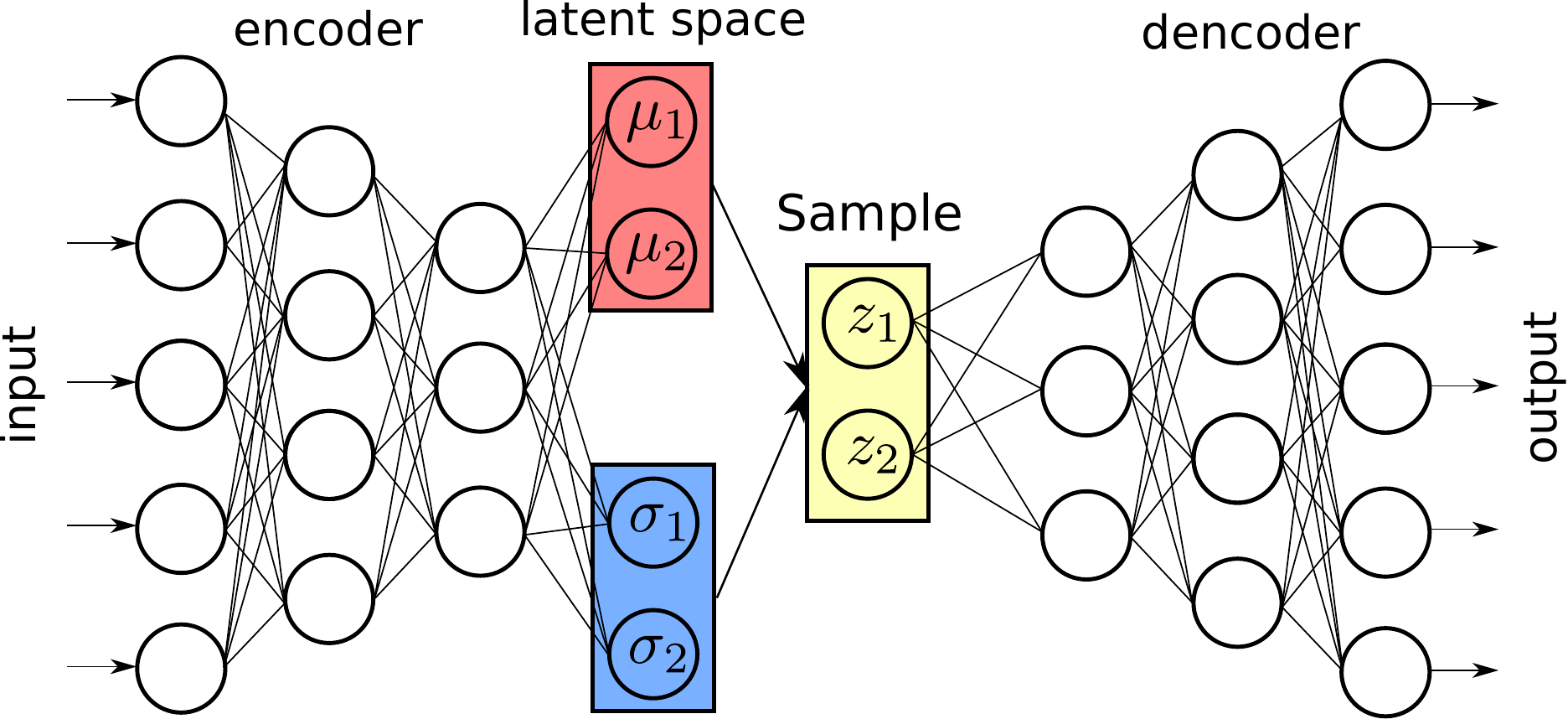}
\caption{Architecture of variational autoencoder (VAE) for unsupervised learning of regular-chaos transition in quantum billiards.}\label{fig:VAE}
\end{figure}

We perform unsupervised (autosupervised) learning of two classes (``regular'' and ``chaotic'') using a variational autoencoder (VAE). 
The unlabeled dataset was prepared in a similar way as for the supervised learning. 
Dataset consist of randomly sampled images of $|\psi_n(x,y)|^2$ with the dimensions $36\times 36$, number of samples in the training dataset for each billiard type is $6\times 10^3$, number of testing samples is $2\times 10^3$ for each billiard type. 
VAE was trained and tested for the states with $n\sim 500$ in Bunimovich and Sinai's billiards, $\lambda=0$ corresponds to the ``regular'' class, $\lambda=0.4$ corresponds to the ``chaotic'' class.
VAE consists of the encoder $Q_{\theta}(x_i)$, decoder $P_{\theta'}(z_i)$, Gaussian sampler $G_{\mu_j,\sigma_j}$ and the latent space of dimension 2 (latent space parameters $\mu_{1,2}$ and $\sigma_{1,2}$) 
representing the two classes, ``regular'' and ``chaotic'', the architecture of VAE is shown in Fig. \ref{fig:VAE}. Here $x_i$ is the vectorized representation of the input data (image), $\theta$ ($\theta'$) are NN parameters of the encoder (decoder).
The sampler generates random latent space variables $z_{1,2}$ with the mean $\mu_{1,2}$ and the dispersion $\sigma_{1,2}^2$.   
Decoder performs reconstruction from the latent space representation to  the original data format, the `image' $|\psi(x,y)|^2$ with dimensions equal to the input dimension ($36\times 36$). 
Final layer of the decoder has sigmoid activation function in order to match the input data range (we normalize the input data so that $\max\{|\psi(x,y)|^2\}=1$.
Encoder and decoder are represented by a fully connected NN with two hidden layers and $N_h=150$ neurons in each layer. 
The structure of the decoder network replicates the structure of the encoder (number of layers, number of neurons, activation function) and the decoder is a `mirrored' replica of the encoder.  
The encoder network is given by two fully connected layers with ReLU activation function between the layers. 

The objective function is a sum of reconstruction loss (binary cross entropy) and KL divergence loss: \cite{Kingma2014}
\begin{equation}\label{eq:L_VAE}
	\mathcal{L}_{VAE}(x) = \mathbb{E}_{z\sim Q_\theta(x)} [ \log P_{\theta'} (z) ] - \frac{1}{2}  \sum_{j=1,2}(1 + \log {\sigma_j^2} - \mu_j^2 - \sigma_j^2),
\end{equation}
where $\mathcal{L}_{VAE}$ is the loss function, $x$ is the data sample (discretized  wave function image $|\psi|^2$), $P_{\theta'}$ is the output of the decoder network. 
Expectation value $\mathbb{E}_{z\sim Q_\theta(x)} [\ldots]$ is evaluated by averaging over batch of $z_{j}$ sampled from the latent space.
Objective function (\ref{eq:L_VAE}) is also known as the variational lower bound or 
Evidence Lower Bound (ELBO). 
This is the bound on the log-probability to observe a data point $x$, therefore by maximizing the lower bound  (\ref{eq:L_VAE}) we maximize the log-likelihood probability of observation $x$.
We implemented VAE within PyTorch framework.
VAE was trained over 50 epochs using Adam optimizer~\cite{Kingma2014a}, learning rate is $\alpha=10^{-4}$, batch size is $40$ samples.

\section{Anomaly detection with VAE and quantum scars}\label{app:anom}

Among the  wave functions of Sinai and Bunimovich billiards we selected states with scarred wave functions. 
The total number of scarred wave functions  constituted only a small fraction of the entire dataset ($<5 \%$). 
Some typical examples of scarred wave functions in Bunimovich and Sinai billiards are shown in the left panel of Fig.~\ref{fig:scars_pics}.
We train VAE on the entire dataset containing chaotic and regular wave functions. 
At test time we feed real space images of wave function snapshots $|\psi_n(x,y)|$ to VAE and analyse the latent space representaion $z_j\sim G_{\mu_j, \sigma_j}(Q_{\theta}(x_i))$ of the input samples $x_i$. 
The portion of ``scarred'' samples in the test dataset is $ 33\%$, such ratio was chosen to make `scarred' clusters in the latent space well visible.

\begin{widetext}
\begin{center}
\begin{figure}[htb!]
\includegraphics[scale=0.25]{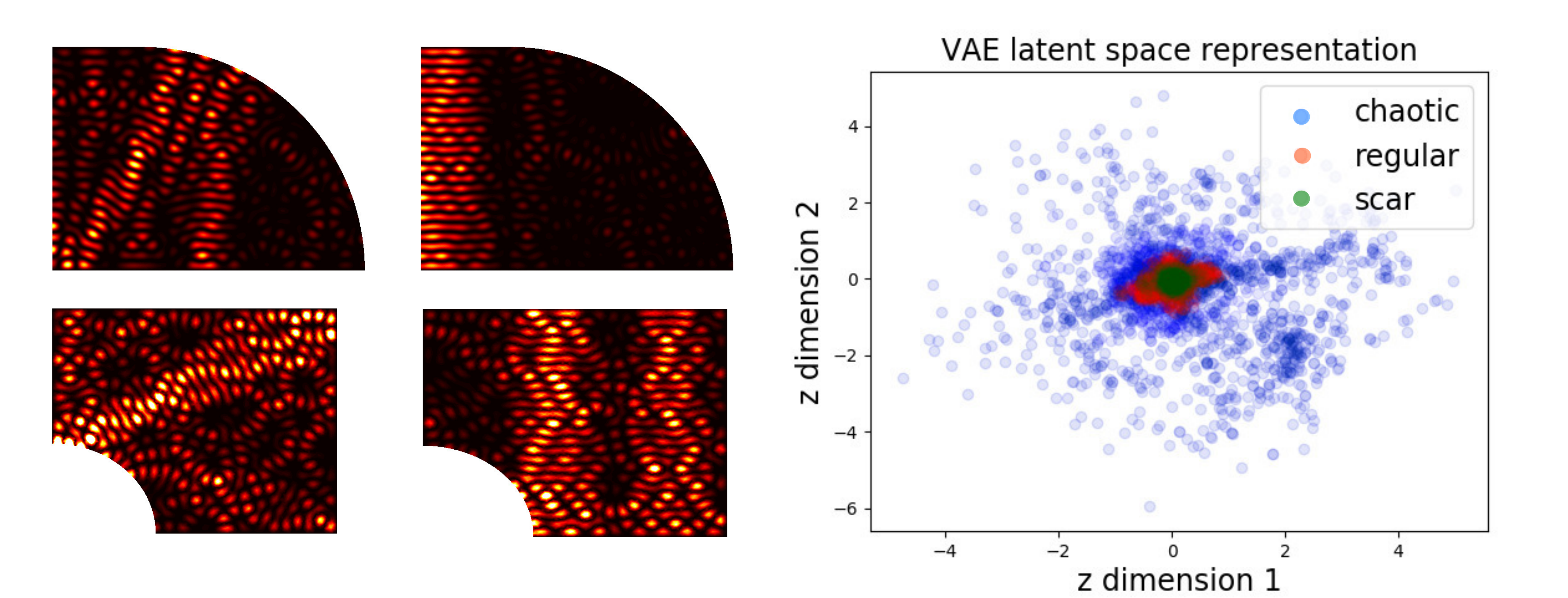}
\caption{Left panel: Examples of scarred wave functions in Bunimovich and Sinai billiards used for quantum anomaly detection.
Right panel: Latent space distribution of VAE for 'regular' ($\lambda=0$) and 'chaotic' ($\lambda=0.4$) wave functons wave functions  in Sinai billiard. Green dots correspond to scared wave functions.}
\label{fig:scars_pics}
\end{figure}
\end{center}
\end{widetext}

Scarred chaotic wave functions form a cluster in the ``wrong'' region that strongly overlaps with the ``regular'' cluster, see Fig. \ref{fig:scars_pics}, right panel and Fig 2(c) in the main text. 
This is a signature of anomalous properties of scarred wave functions that we use as a case for demonstration of the anomaly detection approach.
In Fig. \ref{fig:scars_pics} (right panel) we show how `regular' ($\lambda=0$) ,  `chaotic' ($\lambda=0.4$) and `scarred' wave functions ($\lambda=0.4$) of Sinai billiard are represented in the VAE's latent space. 
Another popular VAE-based approach for anomaly detection relies on the increase of VAE reconstruction loss (or reconstruction probability) of anomalous data~\cite{Zhang2019}.
This approach does not pertain to our case, because the reconstruction loss for scarred wave functions is approximately equal to the reconstruction loss for `regular' wave functions.

In addition to anomaly detection, we use VAE  latent space representation to explore possibility of  smooth interpolation between wave functions corresponding to regular and chaotic billiards. 
By scanning across coordinates in the  latent space  $z_{1}, z_2$ and decoding the latent representation  
with the decoder neural network $y\sim P_{\theta'}(z)$ into vectorized form corresponding to the original data dimensions, we obtained `images' of wave functions $|\psi|^2$ (Fig. 2d, main text) 
interpolating between chaotic (center region)  and regular wave functions (outer region). 

\section{Dataset preparation for Heisenberg XXZ chains (exact diagonalization)} \label{app:data_xxz}

We find eigenstates of Heisenberg XXZ model for an arbitrary value of perturbation parameter $\lambda$ by the exact diagonalization method based on the Lancsoz algorithm~\cite{Sandvik2011}.
We used Python implementation of the QuSpin software package~\cite{Weinberg2017}. 
In order to avoid extensive computational costs, the size of Hamiltonian matrix was reduced by considering only the eigenstates in certain parity and magnetization sectors of the XXZ Heisenberg model. 
Specifically, we find eigenstates in the even parity sector and the lowest magnetization sector. 
The lowest magnetization sector corresponds to the states with $m_{z} = \left({n_{\uparrow}-n_{\downarrow}}\right)/{2} = 1/2$ (for odd spin chains),
where $n_{\uparrow}$ and $n_{\downarrow}$ the number of up and down spins, respectively. 

Dataset for Heisenberg XXZ chains consists of vectors of probability densities (PDs) $|\langle \psi_n | k \rangle|^2$ corresponding to integrable and chaotic Hamiltonians.
We take the wave function $|\psi_n\rangle$ corresponding to a quantum state with the energy lying in the center of the spectrum.
In order to prepare a diverse dataset for a given value of $\lambda$ we randomly select $J_{zz}$ from the uniform distribution $J_{zz} \in [0.8, 2]$.  
Since the XXZ model is integrable for any value of $J_{zz}$  we build a dataset corresponding to a set of different Hamiltonians by varying $J_{zz}$.
In the training set we include PDs for regular systems ($\lambda=0$) and chaotic systems ($\lambda_0=0.3$) and label the samples, accordingly. 
The test set contains PDs corresponding to a discrete set of $\lambda$ lying in the interval $\lambda\in[0,0.3]$. 
The training set contains 400 samples, testing set consists of 100 samples.

\vspace{1cm}
\section{Multi-layer perceptron} \label{app:mlp} 

\begin{figure}[h!]
	\includegraphics[scale=0.7]{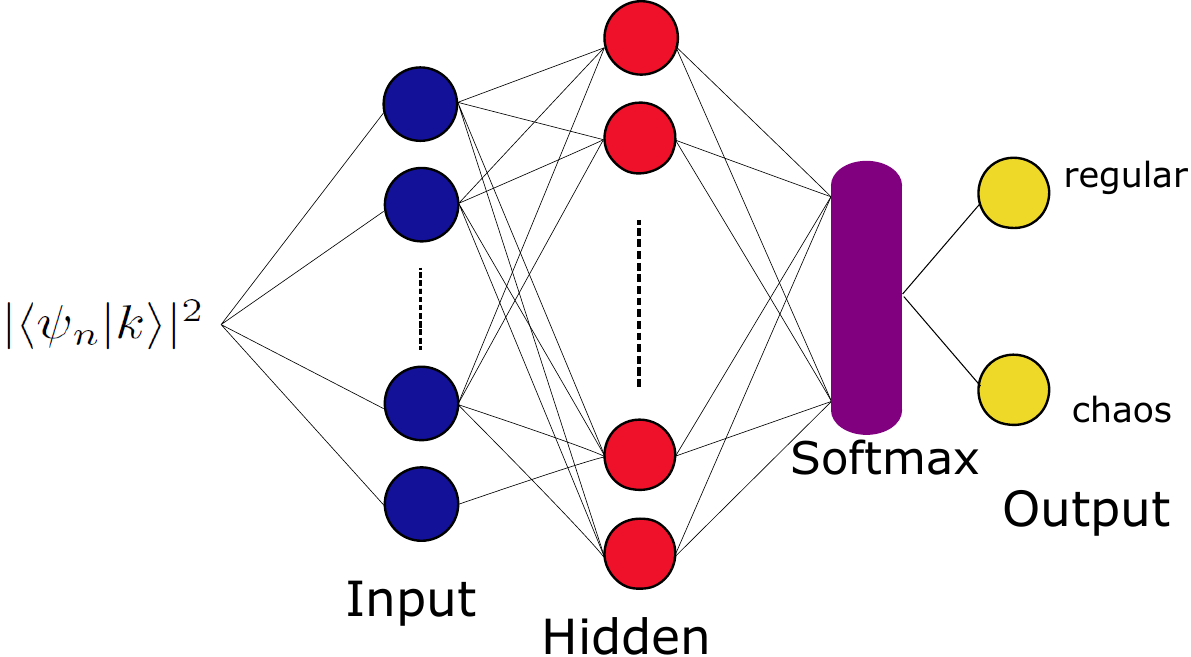}
	\caption{Multilayer perceptron used for investigation integrable/chaotic transitions in Heisenberg XXZ chains.}
	\label{fig:chains_mps}
\end{figure}

\begin{widetext}
\begin{center}
\begin{figure}[h!]
\includegraphics[scale=0.8]{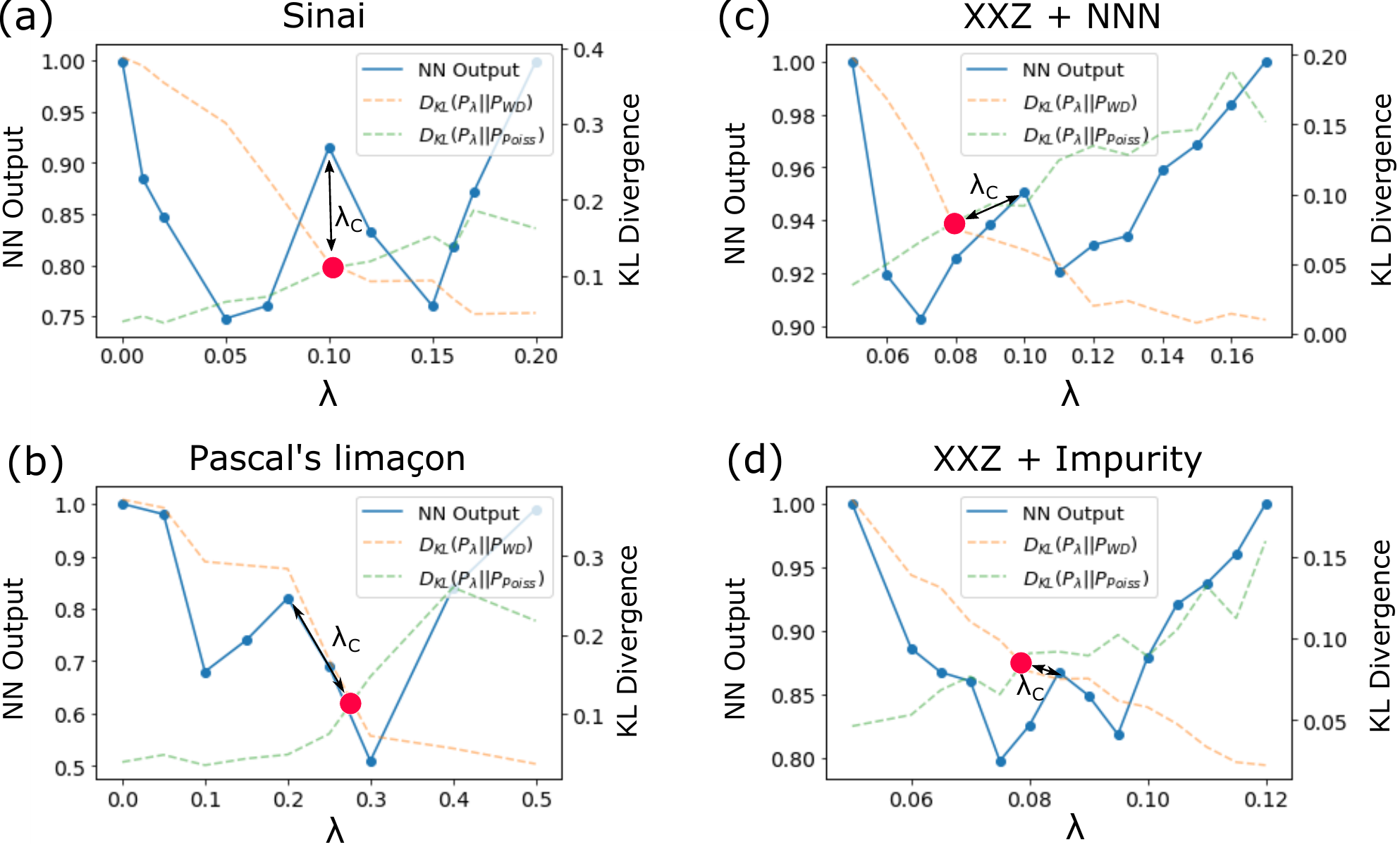}
\caption{Universal W-like NN performance curves in the `learning by confusion'  scheme for the Sinai billiard (a); the Pascal's lima\c{c}on (b); XXZ + NNN (c) and XXZ + Impurity (d). The predicted transition point $\lambda_c$ is highlighted.
The estimated position of the transition point predicted from the KL divergence calculation is shown with a red dot.}
\label{fig:W-Bill}
\end{figure}
\end{center}
\end{widetext}

We use a standard MLP neural network that consists of an input layer with the size $n$, which is equal to the size of vector with PDs in the specified symmetry (parity and  total magnetization) sector of the eigenstates; 
one hidden layer with $m = 700$ neurons, and an output softmax layer.
Neurons of the hidden layer receive input $x_{i}$ and a weight $w_{xi}$ ($i=1...n$) and compute output $y=f(z)$, where $z=\sum_{i=1}^{n}x_{i}w_{xi}$. 
An output of a neuron is computed with a sigmoid activation function $f(z)=1/(1+e^{-z})$. 
Further, each output $y$ with a corresponding weight $w_{yi}$ ($i=1...m$) is passed to two neurons of an output softmax layer, which finally results in a two-component vector $(p_1, p_2)$ that obeys the constraint $p_1+p_2=1$.  
The softmax layer for binary classification task is defined as $p_{j=1,2} = \frac{\exp{y_{j}}}{\sum_{i=1,2} \exp {y_i}}$.
The scalar values $p_1$ ($p_2$) are interpreted as a probability that the input wave function belongs to the regular (chaotic) class.
The objective function is the binary cross-entropy. 
Neural network's weights are optimized using Adam optimizer \cite{Kingma2014a} with the learning rate $\alpha=0.001$, batch size of $10$ samples, $20$ training epochs. 
The scheme of the neural network architecture is presented in Fig.~\ref{fig:chains_mps}.

We used densely connected MLP instead of CNN architecture due to the following reason: CNN is designed to grasp spatial structure of the input data, whereas MLPs are used for more general tasks.
CNN architecture is very natural for  image recognition tasks (in our case -  classifying wave functions in quantum billiards), but generically is  not natural representation for the case of the spin chains, 
where the input data corresponds to the components of the many-body wave function.

\vspace{1cm}

\section{Detection of critical points with a confusion scheme with confusion method (W-shape curves)}\label{app:w}
W-like neural network performance curves versus chaoticity parameter $\lambda$ found by ``learning by confusion'' approach  for quantum billiards and XXZ spin chains are shown in Fig. \ref{fig:W-Bill}. 
The central peak of the W-like curve corresponds to the transition point $\lambda_c$ predicted by the neural network.

\end{document}